\documentstyle[12pt]{article}
\begin{document}
\baselineskip 0.7cm

\newcommand{\gsim}{ \mathop{}_{\textstyle \sim}^{\textstyle >} }
\newcommand{\lsim}{ \mathop{}_{\textstyle \sim}^{\textstyle <} }
\newcommand{\vev}[1]{ \left\langle {#1} \right\rangle }
\newcommand{\lsp}{ \left ( }
\newcommand{\rsp}{ \right ) }
\newcommand{\lmp}{ \left \{ }
\newcommand{\rmp}{ \right \} }
\newcommand{\llp}{ \left [ }
\newcommand{\rlp}{ \right ] }
\newcommand{\labs}{ \left | }
\newcommand{\rabs}{ \right | }
\newcommand{\KEV}{ {\rm } }
\newcommand{\MEV}{ {\rm MeV} }
\newcommand{\GEV}{ {\rm GeV} }
\newcommand{\TEV}{ {\rm TeV} }
\newcommand{\mgut}{M_{GUT}}
\newcommand{\mint}{M_{I}}
\newcommand{\mgra}{M_{3/2}}
\newcommand{\mll}{m_{\tilde{l}L}^{2}}
\newcommand{\mdr}{m_{\tilde{d}R}^{2}}
\newcommand{\mllXX}[1]{m_{\tilde{l}L , {#1}}^{2}}
\newcommand{\mdrXX}[1]{m_{\tilde{d}R , {#1}}^{2}}
\newcommand{\mgy}{m_{G1}}
\newcommand{\mgl}{m_{G2}}
\newcommand{\mgc}{m_{G3}}
\newcommand{\nuR}{\nu_{R}}
\newcommand{\slL}{\tilde{l}_{L}}
\newcommand{\slLi}{\tilde{l}_{Li}}
\newcommand{\sdR}{\tilde{d}_{R}}
\newcommand{\sdRi}{\tilde{d}_{Ri}}

\begin{titlepage}

\begin{flushright}
UT-768\\
Jan., 1997
\end{flushright}

\vskip 0.35cm
\begin{center}
{\large \bf A Solution to the $\mu$ Problem in Gauge-mediated
Supersymmetry-breaking Models }
\vskip 1.2cm
{\bf T.~Yanagida}
\vskip 0.4cm

{\it  Department of Physics, University of Tokyo, Tokyo 113, Japan}

\vskip 1.5cm

\abstract{
We point out that a sector required to set the cosmological constant 
to zero in gauge-mediated supersymmetry-breaking models naturally produces
a supersymmetry-invariant mass ($\mu$ term) for Higgs doublets
of the order of the electroweak scale. Since this new sector preserves the 
supersymmetry, it does not generate supersymmetry-breaking masses for
the Higgs doublets and thus the $\mu$ problem is solved.
}

\end{center}
\end{titlepage}

%
%
%
%
Low-energy supersymmetry (SUSY) breaking with gauge mediation \cite{DN} has,
recently, attracted much  attention in particle physics. This is not only
because it has a number of phenomenological virtures \cite{dine} but also
because various mechanisms for dynamical SUSY breaking have been discovered
\cite{IY}\cite{dsb}.
This approach, however, has several drawbacks. The most serious one is the
so-called $\mu$ problem. That is, if one wants to generate dynamically
a SUSY-invariant mass ($\mu$ term) for the Higgs doublets by introducing their
coupling to the SUSY-breaking sector, one necessarily gets too large SUSY-
breaking mass ( called $B$ term) for the Higgs doublets \cite{mu-problem}.
Various solutions to this problem have been proposed \cite{mu-problem}
\cite{DN}, but no compelling one has been found so far.    

In this short paper we point out that a sector required to set the cosmological
constant to zero in gauge-mediated SUSY-breaking (GMSB) models provides
a natural solution to the $\mu$ problem.

Let us first discuss the cosmological constant problem in the GMSB models.
In these models the SUSY is supposed to
break down dynamicaly through nonperturbative effects of underlying gauge
interactions. Vacuum-expectation values of the SUSY-breaking $F$ term 
and the superpotential $W$ are given by the dynamical scale $\Lambda_{SUSY}$ 
of the underlying gauge interactions as
\begin{eqnarray}
\langle F \rangle \sim \Lambda_{SUSY}^{2}
\end{eqnarray}
and
\begin{eqnarray}
\langle W \rangle \sim \Lambda_{SUSY}^{3}.
\end{eqnarray}
In supergravity the vacuum-energy density is fixed roughly by the 
vacuum-expectation values $\langle F \rangle$ and $\langle W \rangle$ as
\begin{eqnarray}
\langle V \rangle \simeq \langle F \rangle^{2} - \frac1{3M^2}
\langle W \rangle^{2},
\label{energy}
\end{eqnarray}
where $M$ is the gravitational scale $M\cong 2.4\times 10^{18}$ GeV. Since
the dynamical scale $\Lambda_{SUSY}$ is much smaller than the $M$, the two
terms in Eq.(\ref{energy}) never cancel out. Thus, we always have a
positive vacuum-energy density (cosmological constant).

One may solve this problem by adding a constant term to the superpotential
$W$ by hand. However, it breaks a R symmetry. Once one allows explicit
breaking of the R symmetry, there is no reason why the constant term should
be so small compared to the gravitational scale $M^3$. Therefore, it is
quite reasonable to consider that such a constant term in the superpotential
has a dynamical origin. Namely, we consider that the R symmetry is broken
by nonperturbative effects of new gauge interactions so that the induced
constant term in $W$ cancels out the vacuum-energy density given by the 
SUSY-breaking sector. This requires fine tuning, but it seems the least 
assumption in the GMSB models.

We now construct a new sector for the R-symmetry breaking. The simplest
example is a SUSY $SU(2)$ gauge theory with four doublet chiral
superfields $Q_i (1=1,...4)$. We introduce six singlet superfields $S^{ij}
= -S^{ji}$, where $i,j=1,...4$. To make our analysis simpler we impose a
global $SP(2)$ symmetry under which the $Q_i$ transform as {\bf 4} and the
$S^{ij}$ split up as $S_5^{ij}$ and $S_0$ transforming as {\bf 5} and {\bf 1},
respectively. We consider a continuous R symmetry $U(1)_R$ and assume
the R-charges for all matter superfields $Q_i$ and $S^{ij}$ 
are $2/3$. Then a tree-level superpotential is given by\footnote{The model
with $g_5=g_0=0$ has been proposed as an example for the dynamical SUSY
breaking in Ref.\cite{IY}.}
\begin{eqnarray}
W_{tree}= \lambda_5Q_iQ_jS_5^{ij} + \frac{\lambda_0}{2}(Q_1Q_2 + Q_3Q_4)S_0
          -\frac{g_5}{2}(S_5^{ij})^2S_0 - \frac{g_0}{6}S_0^3.
\end{eqnarray}

It is known that the full nonperturbative effects in this theory generate
a low-energy superpotential\cite{seiberg} 
\begin{eqnarray}
W_{dyn}= X(PfM_{ij}-\Lambda^4),
\end{eqnarray}
where $M_{ij}$ are the gauge-invariant degrees of freedom 
\begin{eqnarray}
M_{ij}= -M_{ji}\sim Q_iQ_j
\end{eqnarray}
and $X$ is a Lagrange multiplier field. Then, the full low-energy
superpotential is written in terms of the gauge invariants $M_{ij}$
and $S^{ij}$ as
\begin{eqnarray}
W= X(PfM_{ij}-\Lambda^4) + \lambda_5M^5_{ij}S_5^{ij}
       + \frac{\lambda_0}{2}(M_{12}+M_{34})S_0 - \frac{g_5}{2}(S_5^{ij})^2S_0
       - \frac{g_0}{6}S_0^3.
\end{eqnarray}
Here, $M^5_{ij}$ denotes the {\bf 5} representation of the $SP(2)$ while
$(M_{12}+M_{34})$ transforms as a singlet under the global $SP(2)$.

This model possesses various SUSY-invariant vacca\footnote{In the limit of both
$g_5$ and $g_0$ vanishing all the SUSY-invariant vacua run away to infinity
\cite{IY}.}, among which we choose the following $SP(2)$-invariant one\footnote
{In the $SP(2)$-violating vacua we have Nambu-Goldstone superfields. 
If one introduces explicit breaking of the global $SP(2)$ one can eliminate
such unwanted massless fields. In this case one may choose the $SP(2)$-
violating vacua, which does not, however, change any essential point in this
paper.}:
\begin{eqnarray}
\label{vacuum1}
\langle M_{12} \rangle = \langle M_{34} \rangle = \Lambda^2,\\
\langle S_0 \rangle = \sqrt{\frac{2\lambda_0}{g_0}}\Lambda,\\
\langle M^5_{ij} \rangle = \langle S_5^{ij} \rangle = 0.
\label{vacuum2}
\end{eqnarray}
In this vacuum the continuous R symmetry breaks down to a discrete $Z_{2R}$
, which generates a constant term in the superpotential $W$ as
\begin{eqnarray}
\langle W \rangle = \sqrt{\frac{8\lambda_0^3}{9g_0}}\Lambda^3.
\label{vacuum}
\end{eqnarray}
Notice that there appears no R-axion superfield. This is because the R symmetry
defined at the classical level is broken explicitly by nonperturbative 
effects at the quantum level since the R-current has an $SU(2)$ gauge anomaly.

Since the SUSY is preserved in this sector, the constant superpotential
Eq.(\ref{vacuum}) leads to a negative vacuum-energy density
\begin{eqnarray}
\langle V^{\prime } \rangle \simeq -\frac{1}{3M^2}\langle W \rangle^2,
\end{eqnarray}
which is supposed to cancel the positive vacuum-energy density 
Eq.(\ref{energy}) coming from the SUSY-breaking sector. This gives us 
a relation between the two dynamical scales
$\sqrt{\langle F \rangle} \sim \Lambda_{SUSY} $ and $\Lambda $ as
\begin{eqnarray}
\langle F \rangle ^2 \simeq \frac {1}{3M^2}\langle W \rangle ^2\\
~~~~~~= \frac{8\lambda_0^3}{27g_0M^2}\Lambda^6.
\label{relation}
\end{eqnarray}

We are now at the point to discuss the $\mu$ term for a pair of Higgs doublets 
$H$ and $\overline{H}$. The form of coupling of the Higgs doublets
to this R-symmetry breaking sector depends on the R-charge of the product
$H\overline{H}$. We assume the product $H\overline{H}$ has the R-charge
$= 2/3$. Then, the Higgs doublets have the following nonrenormalizable
couplings to $Q_i$ and $S^{ij}$ in the superpotential:
\begin{eqnarray}
W_{H,\overline{H}} = \frac{h}{2M}(Q_1Q_2 + Q_3Q_4)H\overline{H}  
         + \frac{k}{M}(S_5^{ij})^2H\overline{H} + \frac{k^{\prime}}
{M}S_0^2H\overline{H}.
\label{EQ}
\end{eqnarray}
In the vacuum Eqs.(\ref{vacuum1})-(\ref{vacuum2}) we have a SUSY-invariant
mass $\mu$ for the Higgs doublets as
\begin{eqnarray}
\mu = \frac{h}{2M}(\langle M_{12}\rangle + \langle M_{34}\rangle )
~ =\frac{h}{M}\Lambda^2.
\label{mu1}
\end{eqnarray}
Here, we have neglected the contribution from the third term in Eq.(\ref{EQ})
for simplicity, since it does not change the essential point of our 
conclusion. From Eqs.(\ref{relation}) and (\ref{mu1}) we get
\begin{eqnarray}
\mu \simeq \frac{3hg_0^{1/3}}{2\lambda_0}\left(\frac{\sqrt{F}}{10^6GeV}\right)
^{4/3}\times 100GeV.
\end{eqnarray}
We see that the mass obtained for the Higgs doublets lies
in the electroweak mass region for a reasonable range of parameters
$(hg_0^{1/3}/\lambda_0)\simeq 0.1-10 $ and $\sqrt{F}\simeq 10^5-10^7$ GeV
\cite{DN}\cite{HIY}. Since the R-symmetry breaking sector 
preserves the SUSY, it never generates SUSY-breaking masses for the Higgs 
doublets. Thus, we have no $\mu$ problem.

It is now clear that the gauge invariant $Q_iQ_j$ which condenes in the 
vacuum play a central role in our solution. We may construct various,
similar models for the R-symmetry breaking sector. For example, consider
an $SU(N_C)$ gauge theory with a pair of matter superfields $Q$ and 
$\overline{Q}$ transforming as ${\bf N_C}$ and ${\bf N_C^{*}}$
under the $SU(N_C)$, respectively. Introduction of a singlet $S$ allows
us to have a tree-level superpotential
\begin{eqnarray}
W_{Tree} = \lambda Q\overline{Q}S - \frac{g}{3}S^3.
\end{eqnarray}
In this model ($N_C > N_F =1$) the full nonperterbative effects yield the
exact effective superpotential at low energies as \cite{ADS}
\begin{eqnarray}
W_{eff.} = b\Lambda^{\frac{3N_C-1}{N_C-1}}(Q\overline{Q})^{\frac{-1}{N_C-1}}
          + W_{Tree}.
\end{eqnarray}
We have a SUSY-invariant vacuum
\begin{eqnarray}
\langle Q\overline{Q}\rangle = \left(\frac{g}{\lambda^3}\frac{b^2}{(N_C-1)^2}
\right)
^{\frac{N_C-1}{3N_C-1}} \Lambda^2\\
\langle S \rangle = \frac1{\lambda}\left(\frac{g}{\lambda^3}\right)
^{\frac{-N_C}{3N_C -1}}\left(\frac{b}{N_C-1}\right)^{\frac{N_C-1}{3N_C-1}}
\Lambda.
\end{eqnarray}
As in the previous model a nonrenormalizable interaction $\frac{h}{M}
Q\overline{Q}H\overline{H}$ produces the $\mu$ term for the Higgs doublets
of the order of the electroweak scale.

In conclusion, we should stress that the vanishing cosmological constant
in the GMSB models suggests the existence of new dynamics at the scale
$\Lambda \sim 10^{10}$ GeV. We have no explanation for why the contributions
from the SUSY-breaking and the R-symmetry breaking sectors to the 
vacuum-energy density cancel out exactly, but it seems a
necessary condition for any realistic GMSB model. On the other hand, it 
is very  
encouraging that the new sector required to set the cosmological constant
to zero naturally produces a $\mu$ term for the Higgs doublets in the
desired mass region, $\mu \sim O(100)$ GeV.

\vskip 1.5cm
\noindent{\bf Acknowledgment}

We thank E.~Stewart for useful comments.

\newpage

%
%
\newcommand{\Journal}[4]{{\sl #1} {\bf #2} {(#3)} {#4}}
\newcommand{\PL}{\sl Phys. Lett.}
\newcommand{\PR}{\sl Phys. Rev.}
\newcommand{\PRL}{\sl Phys. Rev. Lett.}
\newcommand{\NP}{\sl Nucl. Phys.}
\newcommand{\ZP}{\sl Z. Phys.}
\newcommand{\PTP}{\sl Prog. Theor. Phys.}
\newcommand{\NC}{\sl Nuovo Cimento}

%

\end{document}